\documentclass[fleqn,usenatbib]{mnras}

\usepackage{newtxtext,newtxmath}

\usepackage[T1]{fontenc}

\DeclareRobustCommand{\VAN}[3]{#2}
\let\VANthebibliography\thebibliography
\def\thebibliography{\DeclareRobustCommand{\VAN}[3]{##3}\VANthebibliography}

\usepackage{graphicx}	
\usepackage{amsmath}	
\usepackage{subcaption}
\usepackage[autostyle, english=british]{csquotes}

\title[Optimising photometric redshift distributions]{Optimising the shape of photometric redshift distributions with clustering cross-correlations}

\author[St\"olzner et al.]{
Benjamin St\"olzner$^{1,2}$\thanks{E-mail: stoelzner@astro.rub.de (BS)},
Benjamin Joachimi$^{1}$,
Andreas Korn$^{1}$,
and the LSST Dark Energy Science Collaboration
\\
$^{1}$Department of Physics and Astronomy, University College London, Gower Street, London WC1E 6BT, UK\\
$^{2}$Ruhr-University Bochum, Astronomical Institute (AIRUB), German Centre for Cosmological Lensing, Universit\"atsstr. 150, 44801, Bochum, Germany
}

\date{Accepted XXX. Received YYY; in original form ZZZ}

\pubyear{2022}

\begin{document}
\label{firstpage}
\pagerange{\pageref{firstpage}--\pageref{lastpage}}
\maketitle

\begin{abstract}

We present an optimisation method for the assignment of photometric galaxies into a chosen set of redshift bins. This is achieved by combining simulated annealing, an optimisation algorithm inspired by solid-state physics, with an unsupervised machine learning method, a self-organising map (SOM) of the observed colours of galaxies. Starting with a sample of galaxies that is divided into redshift bins based on a photometric redshift point estimate, the simulated annealing algorithm repeatedly reassigns SOM-selected subsamples of galaxies, which are close in colour, to alternative redshift bins. We optimise the clustering cross-correlation signal between photometric galaxies and a reference sample of galaxies with well-calibrated redshifts. Depending on the effect on the clustering signal, the reassignment is either accepted or rejected. By dynamically increasing the resolution of the SOM, the algorithm eventually converges to a solution that minimises the number of mismatched galaxies in each tomographic redshift bin and thus improves the compactness of their corresponding redshift distribution. This method is demonstrated on the synthetic LSST cosmoDC2 catalogue. We find a significant decrease in the fraction of catastrophic outliers in the redshift distribution in all tomographic bins, most notably in the highest redshift bin with a decrease in the outlier fraction from 57 per cent to 16 per cent.

\end{abstract}




\section{Introduction}

The calibration of the redshift distribution of cosmological surveys plays a crucial role in current studies of cosmology. While spectroscopic observations of galaxies allow for accurate redshift measurements of the source redshift distribution, complete spectroscopic measurements are often infeasible given the large number of observed objects in current and upcoming surveys, such as the Vera C. Rubin Observatory's Legacy Survey of Space and Time \citep[LSST;][]{LSST} and the European Space Agency's Euclid survey \citep{Euclid}. Therefore, surveys often rely on multi-band photometry to determine the redshift of observed objects \citep[see][for a review]{Salvato19}. However, photometric methods suffer from systematic biases and catastrophic outliers in the redshift estimation and thus require a sophisticated calibration of the redshift distribution in order to derive robust constraints on cosmology \citep[see for example:][]{Ma06, Huterer06, Bernstein10, Cunha14}.

Cosmological analyses, for example studies of weak gravitational lensing by the large-scale structure of the Universe, are often performed tomographically, which allows for the utilisation of information about the evolution of the Universe. In tomographic cosmic shear analyses the galaxy sample is split into several redshift bins using photometric redshift estimates of individual galaxies. The cosmic shear signal is then estimated by measuring the cross-correlation between the shapes of galaxies in the tomographic bins, which improves constraints on cosmological parameters \citep{Hu99}. 

A tomographic analysis usually requires two steps. First, the sample of galaxies needs to be divided into redshift bins. This is usually done using galaxy photometry, which are used to estimate the redshift of individual galaxies in the survey, for example via spectral energy distribution (SED) template fitting codes. However, the true redshift distributions of tomographic bins extend beyond the bin boundaries because of noise, systematic biases and catastrophic outliers in the photometric redshift estimation. Therefore, the second step is the calibration of the actual redshift distribution of each tomographic bin which is important when deriving theoretical predictions for the observed weak lensing signal given the sensitivity of the observed signal to the tails of the redshift distribution \citep{Ma06}. For example, such calibration methods include angular cross-correlation clustering measurements with overlapping spectroscopic reference samples \cite[e.g.][]{Newman08,Matthews10, Menard13, McQuinn13, Mcleod17, vdBusch20, Gatti21} and direct calibration methods with spectroscopic subsamples that are, potentially after re-weighting, representative of the full sample \citep{Lima08, Bonnett16, Hildebrandt17}. Furthermore, hierarchical Bayesian models that combine photometry measurements of individual galaxies and clustering measurements with tracer populations in a robust way have been used for redshift calibration \citep{Sanchez19, Alarcon20}. Additionally, the clustering properties of photometric galaxies can be utilised to increase the precision of photometric redshifts \citep{Jasche12}. Moreover, self-organising maps (SOMs) can be used to assign galaxies to groups based on their photometry \citep{Masters15, Buchs19, Wright20, Myles21}, which allows one to derive subsamples of galaxies that are fully represented by spectroscopic reference samples.

In this work we develop a calibration method that improves the first step by reducing the number of outliers in tomographic redshift bins. We develop a method that updates the assigned redshift bin of galaxies in a given photometric catalogue that otherwise would be assigned to an incorrect redshift bins if point estimates of the photometric redshift are used to assign galaxies to bins. The goal is to obtain a sample of galaxies that is divided into well-localised redshift bins. This is achieved by combining a self-organising map, which is used to group galaxies of a similar colour into cells, with measurements of clustering cross-correlations. We make use of point estimates of the photometric redshifts of galaxies to divide a galaxy catalogue into tomographic bins and apply a simulated annealing algorithm to reassign (SOM-based) cells of galaxies to alternative redshift bins. The optimisation algorithm utilises measurements of the clustering cross-correlation between the photometric sample and a reference sample with well-calibrated redshift measurements and maximises correlations between photometric and reference bins of the same redshift while minimising correlations between bins that are disjoint in redshift. We demonstrate the method on the synthetic LSST cosmoDC2 catalogue \citep{Korytov19}.

The paper is structured as follows: The methods that we use are described in Section 2. We show our results in Section 3 and finally we discuss our main conclusions in Section 4.

\section{Methodology}
\begin{figure*}
\centering
\includegraphics[width=\linewidth]{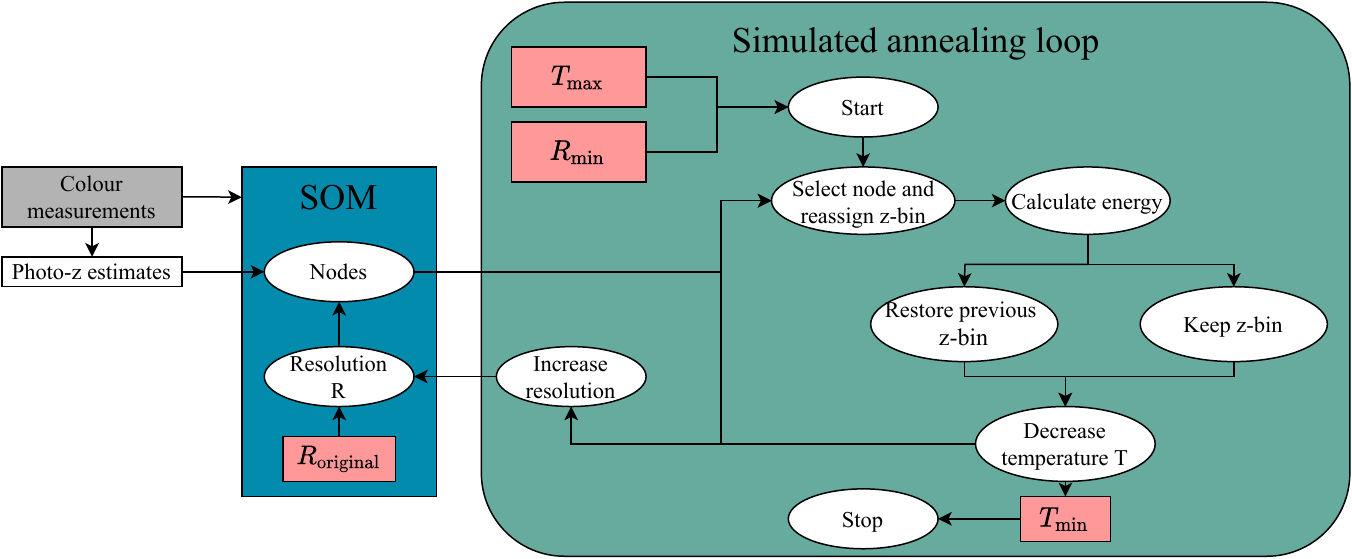}
\caption{Sketch of the optimisation algorithm that reassigns photometric galaxies to alternative redshift bins. We train a self-organising map (SOM) with a high resolution $R_{\rm original}$ on the observed colours of galaxies in the photometric sample, from which we infer SOMs with arbitrary resolutions $R<R_{\rm original}$ . Additionally, we infer point estimates of the photometric redshift to divide the sample into tomographic bins, so that each SOM node is assigned initially to the most common redshift bin of galaxies in this node. We initialise the simulated annealing algorithm with a starting temperature $T_{\rm max}$ and a resolution $R_{\rm min}$, which is coupled linearly to the temperature. In each iteration of the annealing algorithm we select a node of a SOM with the current resolution, which we randomly reassign to a different redshift bin. Measuring the angular cross-correlation between the photometric sample and the reference sample, we calculate the energy of the system from Eq. \eqref{eq:energy}. Comparing the change in energy and the current temperature, we determine whether to keep the redshift bin assignment or to restore the previous state using Eq. \eqref{eq:acceptance}. We then decrease the system's temperature and, depending on the temperature, we either keep the current SOM resolution or increase the resolution using the scheme outlined in Fig. \ref{fig:clustering}. We iterate through these steps until we reach the given final temperature $T_{\rm min}$ and final resolution $R_{\rm max}$.}
\label{fig:algorithm}
\end{figure*}

In this Section we summarise the optimisation method, called {\sc SharpZ}, that we use to assign photometric galaxies to tomographic redshift bins, which is illustrated in Fig. \ref{fig:algorithm}. We start from a catalogue of galaxies that are observed in several photometric bands from which point estimates of the photometric redshift of individual galaxies are estimated. These redshift estimates are used to divide the catalogue into tomographic bins. We aim to minimise mismatches between the photometric redshift and the true redshift of the catalogue that are caused by imprecise redshift estimates in order to infer tomographic bins that are well-localised within the bin boundaries. We employ an overlapping sample of reference galaxies which is divided into the same tomographic bins using accurate redshift measurements in order to quantify how well the true redshift distribution of the photometric sample resembles the redshift distribution of the well-calibrated reference sample. To do so, we measure the angular cross-correlation between the photometric sample and the reference sample. This measurement relies on the property that galaxies cluster spatially, so that we expect a clustering signal between two overlapping photometric and reference samples, whereas samples that are separated in redshift are expected to show no clustering signal. Further details on the clustering measurements can be found in Section \ref{sec:correlation}. We employ a simulated annealing algorithm, explained in Section \ref{sec:annealing}, to randomly reassign photometric galaxies to a different redshift bin in order to maximise the correlation between overlapping photometric and reference bins while minimising the correlation of bins with no overlap in redshift. However, a reassignment of single galaxies only has a marginal impact on the correlation signal between different samples, since the cross-correlation is a statistical property of large samples of galaxies. Therefore, we additionally employ a self-organising map, which is described in Section \ref{sec:SOM}, to derive sets of galaxies of similar colour. By reassigning a set of galaxies in each step of the simulated annealing algorithm, we achieve a measurable effect on the clustering signal between photometric and reference samples, which allows us to employ the combined clustering signals as an objective function to be maximised by the algorithm.

\subsection{Galaxy clustering}
\label{sec:correlation}

In order to determine how well localised within bin boundaries the true redshift distribution of the tomographic bins is, we employ an additional data set comprised of galaxies for which an accurate redshift measurement is available. This can be obtained for example, through spectroscopic observations of galaxies on the same area. Thus, we distinguish between two galaxy samples: 
\begin{enumerate}
\item A photometric galaxy sample of galaxies, which is comprised of objects that are observed through several optical filters. The photometric measurements of those objects are used to infer redshift estimates via the template fitting code {\sc BPZ} \citep{Benitez00}.
\item A reference sample, which is comprised of objects with precise redshift measurements, for example through spectroscopic observations.
\end{enumerate}
Here, we assume for simplicity that the reference sample is fully representative of the photometric sample. While the method can be applied with a reference sample that only partially overlaps with the photometric sample, as discussed in Section \ref{sec:conclusions}, the impact of an inhomogeneous reference sample will be explored in forthcoming work.

Both the photometric sample and the reference sample are divided into $N_{\rm bins}$ redshift bins based on photometric redshift estimates and spectroscopic redshift measurements, respectively. We then measure the two-point correlation function between photometric bins and reference bins on fixed angular scales using the public code {\sc TreeCorr} \citep{Jarvis04}. Using the angular positions of galaxies, we compute the cross-correlation $w^{\rm phot-ref}_{ij}$ between photometric bin $i$ and reference bin $j$ via the Landy-Szalay estimator \citep{Landy93}, which for each bin of angular separation $\Theta$ is defined as
\begin{equation}
\label{eq:ls}
w^{\rm phot-ref}_{ij} = \frac{D^{\rm phot}_iD^{\rm ref}_j - D^{\rm phot}_i R^{\rm ref}_j- R^{\rm phot}_iD^{\rm ref}_j + R^{\rm phot}_i R^{\rm ref}_j}{R^{\rm phot}_i R^{\rm ref}_j},
\end{equation} 
where $D^{\rm phot}_iD^{\rm ref}_j$ denotes the number of observed galaxy pairs of the photometric and reference bins within a single angular bin with range $\theta \in [0.01^{\circ}, 0.1^{\circ}]$. $D^{\rm phot}_i R^{\rm ref}_j$ and $ R^{\rm phot}_iD^{\rm ref}_j$ denote the number of observed galaxy pairs of a random sample with uniform density that follows the geometry of the survey and the photometric and reference bins, respectively. Finally, $R^{\rm phot}_i R^{\rm ref}_j$ denotes the number of galaxy pairs of random samples. 

After calculating the cross-correlation between all combinations of photometric and reference bins, we construct the cross-correlation matrix whose elements are defined via
\begin{equation}
\label{eq:CC_matrix}
\rho_{ij} = \frac{w^{\rm phot-ref}_{ij}}{w^{\rm ref}_j}.
\end{equation}
Here, $w^{\rm ref}_j$ denotes the auto-correlation of reference bin $j$, which serves as a normalisation factor and is calculated by replacing $D^{\rm phot}_i$  with the reference sample $D^{\rm ref}_j$ in Eq.\eqref{eq:ls}.
 
The correlation matrix defined in Eq. \eqref{eq:CC_matrix} acts as a measure of how well the galaxies in each bin of the photometric sample match the underlying true redshift bin. If the redshifts of photometric galaxies were perfectly determined, the correlation matrix between photometric and reference bins would therefore resemble a diagonal matrix. However, we expect non-zero correlation signals between neighbouring redshift bins that are induced by the large structure at their common boundary. The relative magnitude of these off-diagonal correlation signals is dependent on the width of the redshift bins and therefore we expect these signals to be small given the relatively broad redshift bins considered in this work. Additionally, noise and catastrophic outliers in the redshift estimation lead to mismatches between the photometric redshift estimates and the underlying truth and therefore reduce the correlation signal on the diagonal elements. Consequently, they induce a correlation signal on the off-diagonal elements of the cross-correlation matrix. Therefore, we aim to optimise the correlation matrix with the goal to achieve convergence towards a diagonal matrix, which would indicate an optimal assignment of photometric galaxies to redshift bins. The optimisation algorithm requires an objective function, which we define as the difference between the average elements on the diagonal and the off-diagonal elements of the covariance matrix: 
\begin{equation}
E\equiv\frac{1}{N_{\rm bins}}\sum_i \left( \rho_{ii}-\frac{1}{N_{\rm bins}-1}\sum_{i \neq j}\rho_{ij}\right),
\label{eq:energy}
\end{equation}
where $N_{\rm bins}$ denotes the number of tomographic redshift bins. This equation, which quantifies the diagonality of the matrix, defines the so-called \enquote*{energy} of the system, which the simulated annealing algorithm maximises in order to optimise the assignment of photometric galaxies into redshift bins. Furthermore, our choice of normalisation ensures that the energy is independent of the total number of tomographic redshift bins.

Future applications of this work include studies of a more realistic setup where the reference sample consist of a collection of spectroscopically observed galaxies which are not necessarily representative of the photometric sample. In this case, the correlation matrix is dependent on the galaxy bias of the photometric and reference samples. Assuming a linear bias model \citep{Kaiser84}, the mean galaxy overdensity is related to the mean matter overdensity via
\begin{equation}
\delta_{\rm g} = b\,\delta_{\rm m},
\end{equation}
where the bias $b$ can depend on the scale and on colour, redshift, and morphology of galaxies \citep{Fry96, Mann98, Tegmark98}. For a representative reference sample the cross-correlation between the photometric sample and the reference sample is proportional to the product of the biases \citep[see for example][]{Moessner98},
\begin{equation}
w^{\rm phot-ref}_{ij}\propto b^{\rm phot}_ib^{\rm ref}_j, 
\end{equation}
while for the auto-correlation of the reference sample we find
\begin{equation}
w^{\rm ref}_i \propto \left(b^{\rm ref}_i\right)^2.
\end{equation}
Here, we assumed a redshift-dependent galaxy bias, which, however, is assumed to be constant within the boundaries of the photometric redshift bins. The diagonal elements of Eq. \eqref{eq:CC_matrix} for a photometric sample with perfect redshift estimates become
\begin{equation}
\rho_{ii} = \frac{b_i^{\rm phot}}{b_i^{\rm ref}},
\label{eq:identity}
\end{equation}
which is equal to one since we assumed a representative reference sample.

\subsection{Simulated annealing}
\label{sec:annealing} 

To optimise the assignment of photometric galaxies to tomographic redshift bins we employ a simulated annealing algorithm \citep{Kirkpatrick83, Cerny85}, which is a technique inspired by the process of heating and cooling metals to reduce their defects and thus maximising the energy of the given system. This method is applied in optimisation problems in large discrete parameter spaces. In contrast to typical optimisation methods, which usually aim to find the exact optimum, the simulated annealing algorithm achieves an approximation of the global optimum \citep{Mitra86}. In this work, the system is characterised by a set of labels that refer to the redshift bin of each individual galaxy in the photometric sample and the energy of the system is defined in Eq. \eqref{eq:energy}. Given the large number of observed galaxies in photometric surveys we deem exact optimisation methods computationally infeasible and thus we employ the simulated annealing algorithm to optimise the sorting of galaxies into redshift bins. Additionally, the algorithm features a method of avoiding local extrema which allows for finding an approximation of the global optimum of the objective function.

The simulated annealing algorithm works as follows:
The system is characterised by a set of labels $l_k$ that denote the redshift bin to which each photometric galaxy $k$ is assigned. For a given set of labels we use Eq. \eqref{eq:energy} to measure the current energy of the system. Additionally, the system is assigned a temperature $T$ which is a hyperparameter that decreases exponentially from an initial temperature $T_{\rm max}$ to a temperature $T_{\rm min}$. At each iteration of the algorithm the state of the system is altered, i.e. a subset of galaxies is assigned to a different redshift bin, resulting in a change in the system's energy. We then calculate the change in energy $\Delta E$ which is used in conjunction with the temperature to determine whether the altered state is accepted or rejected. A value $\Delta E>0$ indicates that the altered state provides a better solution to the optimisation problem and therefore the new state is accepted. If $\Delta E<0$, the altered state provides a worse solution to the optimisation problem. However, the algorithm allows for a temporary acceptance of a worse solution in order to be capable of leaving local maxima of the objective function and finding the global solution to the optimisation problem. This is achieved by drawing a random number $\alpha$ in the interval $[0,1]$ and comparing the change in energy to the current temperature of the system by evaluating
\begin{equation}
P = \exp\left(\frac{\Delta E}{T}\right).
\label{eq:acceptance}
\end{equation}
If $P > \alpha$, the altered state is accepted and otherwise it is rejected. This allows the algorithm to temporarily explore regions of lower energy that provide a worse solution to the optimisation problem. Since the temperature decreases exponentially, the acceptance probability of a state that worsens the optimisation also decreases over time, so that eventually the algorithm with a high probability only accepts states that provide a better solution to the optimisation problem. Therefore it is important to determine the appropriate setting of the initial and final temperatures, $T_{\rm max}$ and $T_{\rm min}$, so that the algorithm starts with a reasonable probability of accepting worse solutions and finishes at a temperature at which only states that provide a better solution are accepted. 

\subsection{Self-organising maps}
\label{sec:SOM}
The selection of galaxies that are reassigned to a different redshift bin in each iteration is a crucial step in the simulated annealing algorithm. The energy defined in Eq. \eqref{eq:energy} is dependent on the angular cross-correlation between the photometric and reference sample which we optimise by reassigning galaxies to alternative redshift bins. Thus, it is essential to select a set of galaxies that are reassigned together in order to achieve a measurable effect on the objective function. Additionally, we want to select groups of galaxies that are likely to belong to the same tomographic bin, so that they can be reassigned to a common redshift bin. As a tool to select groups of galaxies we use a self-organising map that is trained on the colour measurements of individual galaxies in the photometric sample. This allows for the selection of galaxies of a similar colour, which we expect to also be close in redshift.

A self-organising map \citep[SOM;][]{Kohonen90} is a type of artificial neural network that produces a low-dimensional representation of high-dimensional data using an unsupervised learning technique. In this work, we project a data set containing five colour measurements (u-g, g-r, r-i, i-z, and z-y) onto a two-dimensional space. The map space of the SOM consists of nodes that are arranged on a two-dimensional grid that is usually connected via a rectangular or hexagonal geometry. Furthermore, the topology of the map can be chosen as either planar or toroidal, where the top and bottom as well as the left and right edges are connected to avoid boundary effects. The total number of nodes determines the so-called resolution of the SOM, which determines how well the SOM can separate features in the original data space. In this work, we refer to a map consisting of a rectangular grid of (R x R) nodes as a map of resolution R. For every SOM node there exists a weight vector that links the node to a point in the original high-dimensional data space and thus consists of the corresponding colour values. The training process iteratively alters the randomly initialised weight vectors so that they provide a mapping between the SOM nodes and the data set. In each step the Euclidean distance between the weight vectors and a randomly selected data point is computed. The weight vector of the node that is closest to the data point is called the best matching unit (BMU). Additionally, the neighbourhood of the BMU is identified, which consists of all nodes within a given radius around the BMU. All weight vectors inside the neighbourhood are shifted towards the data point by a fraction of their distance to the data point. This fraction is dependent on the distance between weight vector and data point so that the closer a node is to the BMU, the more its weight vector is shifted. This process is repeated for all data points in the training sample. Moreover, the radius of the neighbourhood around the BMU shrinks over time, so that the number of altered weight vectors in each training step also decreases.  

After training, the weight vectors provide a mapping of the galaxy sample onto a two-dimensional space where galaxies of similar colour are mapped close together while dissimilar galaxies are mapped further apart. Galaxies that are mapped onto a specific node then form a set of galaxies which are close in the original colour space. The total number of SOM nodes then dictates how accurate galaxy clusters in the original colour space can be separated. 

Since the SOM groups galaxies in cells that are similar in colour space, we expect that these galaxies are also close in redshift. Therefore, we make use of the SOM nodes to select sets of galaxies that are assigned to a different redshift bin in each step of the simulated annealing algorithm. Thus, the resolution of the SOM determines the number of galaxies that are relabelled at a time, which imposes a limit on the accuracy of the resulting final assignment of galaxies to redshift bins. While a low-resolution SOM allows us to relabel more galaxies at a time and thus results in a shorter runtime of the algorithm, a high-resolution SOM gives a more accurate result since it allows for a finer separation of galaxies in colour space. Thus, it is advantageous to vary the resolution while running the algorithm, starting with a SOM at a low resolution, denoted $R_{\rm min}$, and increasing the resolution over time up to the maximum resolution $R_{\rm max}$. The advantage of this method is that in the beginning, when we expect the fraction of mislabelled galaxies to be highest, we reassign a larger number of galaxies at a time. By increasing the resolution over time we continuously split the SOM nodes into two, which allows for a finer separation of galaxies in colour space so that the accuracy of the final assignment improves. Finally, we stop the algorithm at a resolution $R_{\rm max}$, at which the average number of galaxies per node becomes so small that continuing the relabelling becomes computationally infeasible given the small impact on the energy of the system.

\begin{figure}
\centering
\includegraphics[width=\linewidth]{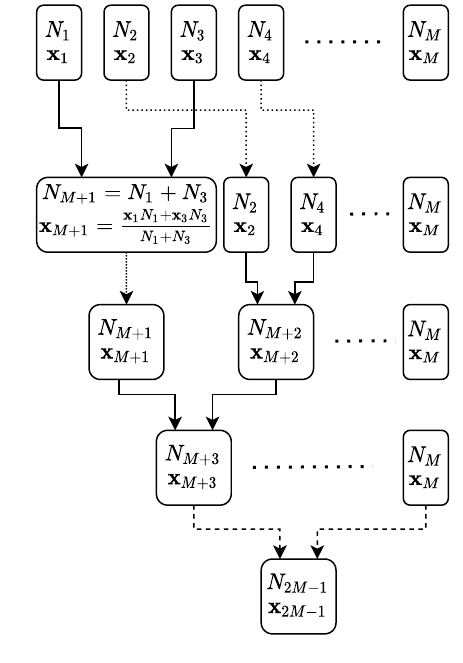}
\caption{Sketch of the clustering method that is used to decrease the resolution of a SOM. We combine nodes of a high-resolution SOM, consisting of $M$ nodes, using the number $N$ of galaxies that are mapped onto each SOM node and the weight vectors $\boldsymbol{x}$, that connect the two-dimensional SOM data space to the original high dimensional data space. In each step we merge the two SOM nodes with the minimum distance of their corresponding weight vectors and compute the weight vector of the combined node as the average of the two weight vectors, where the number of galaxies in each node act as weights. By iterating the merging process we build a hierarchy of clusters until all SOM nodes are merged into one single node after $M-1$ steps. From the hierarchy of nodes we can derive SOMs with any resolutions lower than the resolution of the original SOM.}
\label{fig:clustering}
\end{figure}

In order to be able to dynamically scale the resolution of the SOM, we train a SOM at a high resolution and apply clustering methods to merge nodes that are close in the original data space. From the hierarchy of merged SOM nodes we can then extract a SOM with a lower number of nodes, which corresponds to a lower resolution. This allows us generate SOMs with any arbitrary resolution lower than the original resolution without training multiple SOMs. Standard hierarchical clustering techniques \citep[see for example][]{muellner11} allow for building a hierarchy, where two objects that minimize a given agglomeration criterion are clustered in each step. Thus, this method can be used to iteratively merge the two SOM nodes with the minimum distance between their corresponding weight vectors. However, such clustering methods, applied to the weight vectors of the SOM, do not make use of the information on the number of galaxies in each node. Therefore, we perform the SOM clustering using a weighted method, illustrated in Fig. \ref{fig:clustering}, that utilises the additional information on the number of galaxies per node.

We construct clusters in a bottom-up approach by iteratively merging nodes until only a single node is left. Initially, every node of the high-resolution SOM forms its own cluster. We identify the two nodes with the smallest distance between their corresponding weight vectors. Using the number of galaxies that are assigned to each node as weights, we compute the weight vector of the combined node by calculating the weighted average of the two weight vectors. Starting with a SOM consisting of $M$ nodes, we are left with one single node after $M-1$ clustering steps. After building the hierarchy of merged SOM nodes we can then infer SOMs at resolutions lower than the resolution of the initial SOM.

\section{Results}
\label{sec:results}

We apply the redshift calibration method to the cosmoDC2 catalogue \citep{Korytov19}. This is a large synthetic catalogue designed by the LSST Dark Energy Science Collaboration to support the development of analysis pipelines. In particular, we employ a subset of the {\tt cosmoDC2\_1.1.4} catalogue, covering about 58 ${\rm deg}^2$ of the sky with a magnitude limit of $i<25.3$, which corresponds to the LSST gold sample selection for weak lensing \citep{LSST09}. This catalogue provides colour measurements of approximately $10^7$ galaxies with redshifts $0<z<3$ in the six LSST filter bands (u,g,r,i,z, and y). The photometric redshift is estimated via the template fitting code {\sc BPZ} \citep{Benitez00}. In Appendix \ref{ap:photo-z} we provide a comparison between the point estimate of the photometric redshift and the true redshift of galaxies in the catalogue. Based on the photometric redshift estimate we divide the catalogue into ten bins of equal redshift width between $0<z<2$, where the redshift range of the i-th bin is defined as $\left[0.2(i-1), 0.2i\right]$, and one additional bin with $z>2$. To generate the reference sample of galaxies with well-calibrated redshifts, we draw a random subset of the catalogue that contains 10 per cent of the total galaxies. We assume that we are provided with precise redshift measurements for the galaxies in this subsample and therefore we divide the reference sample into the aforementioned redshift bins using the true simulated redshift. Since in this case the reference sample is perfectly representative of the photometric sample, Eq. \eqref{eq:identity} shows that for an optimal assignment of photometric galaxies the correlation matrix should become close to an identity matrix with small contributions on off-diagonal elements between neighbouring bins, which are induced by the large-scale structure at their common boundary. However, the resolution of the SOM determines the number of galaxies that are typically grouped into one node. Therefore, it imposes a limit on how well the SOM can separate galaxies by redshift, so that we do not expect the correlation matrix to converge to an exact identity matrix.

\begin{figure*}
    \centering
    \begin{subfigure}[t]{0.49\textwidth}
        \centering
        \includegraphics[width=\linewidth]{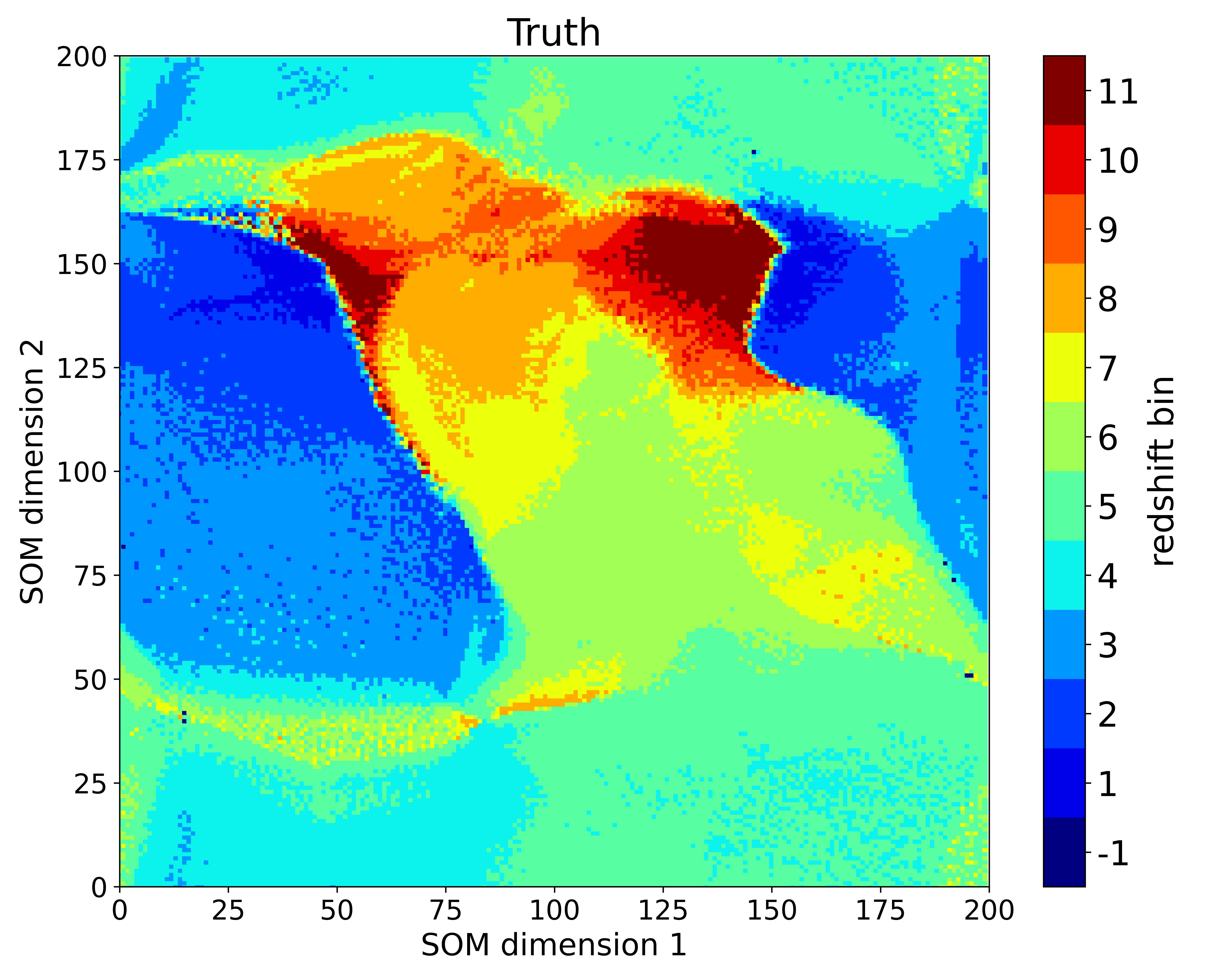} 
        \caption{Colours indicate the photometric redshift bin of each SOM node, based on the true redshift of individual galaxies, which does not enter the optimisation process and is used for illustration purpose of the ideal bin assignment only.}
        \label{fig:SOM}
    \end{subfigure}
    \hfill
    \begin{subfigure}[t]{0.49\textwidth}
        \centering
        \includegraphics[width=\linewidth]{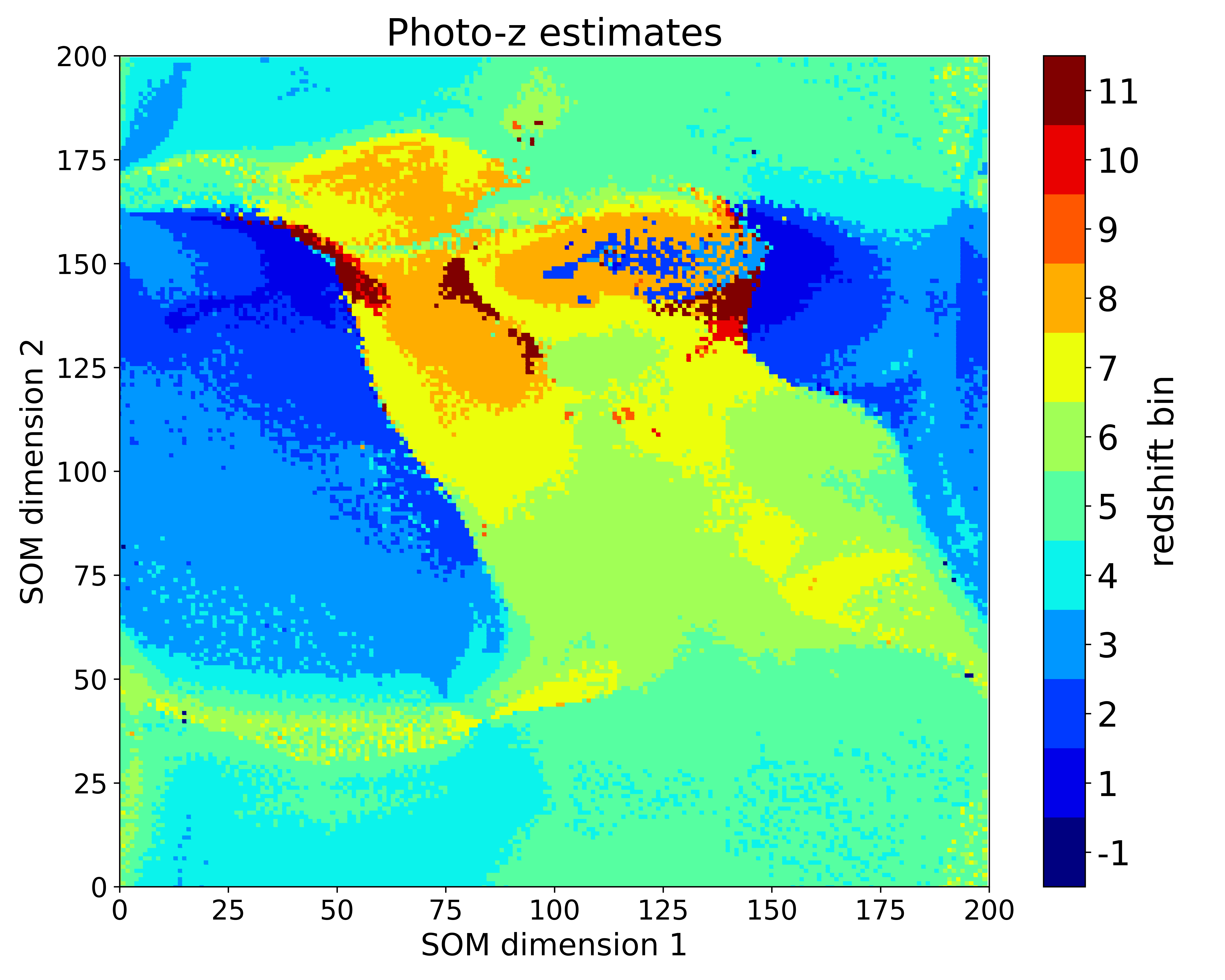} 
        \caption{Colours indicate the initial photometric redshift bins of each SOM node, based on the estimate of the photometric redshift of individual galaxies inferred with \textsc{BPZ}.}
	   \label{fig:SOM_initial_binned}
    \end{subfigure}
    \begin{subfigure}[t]{0.49\textwidth}
        \centering
        \includegraphics[width=\linewidth]{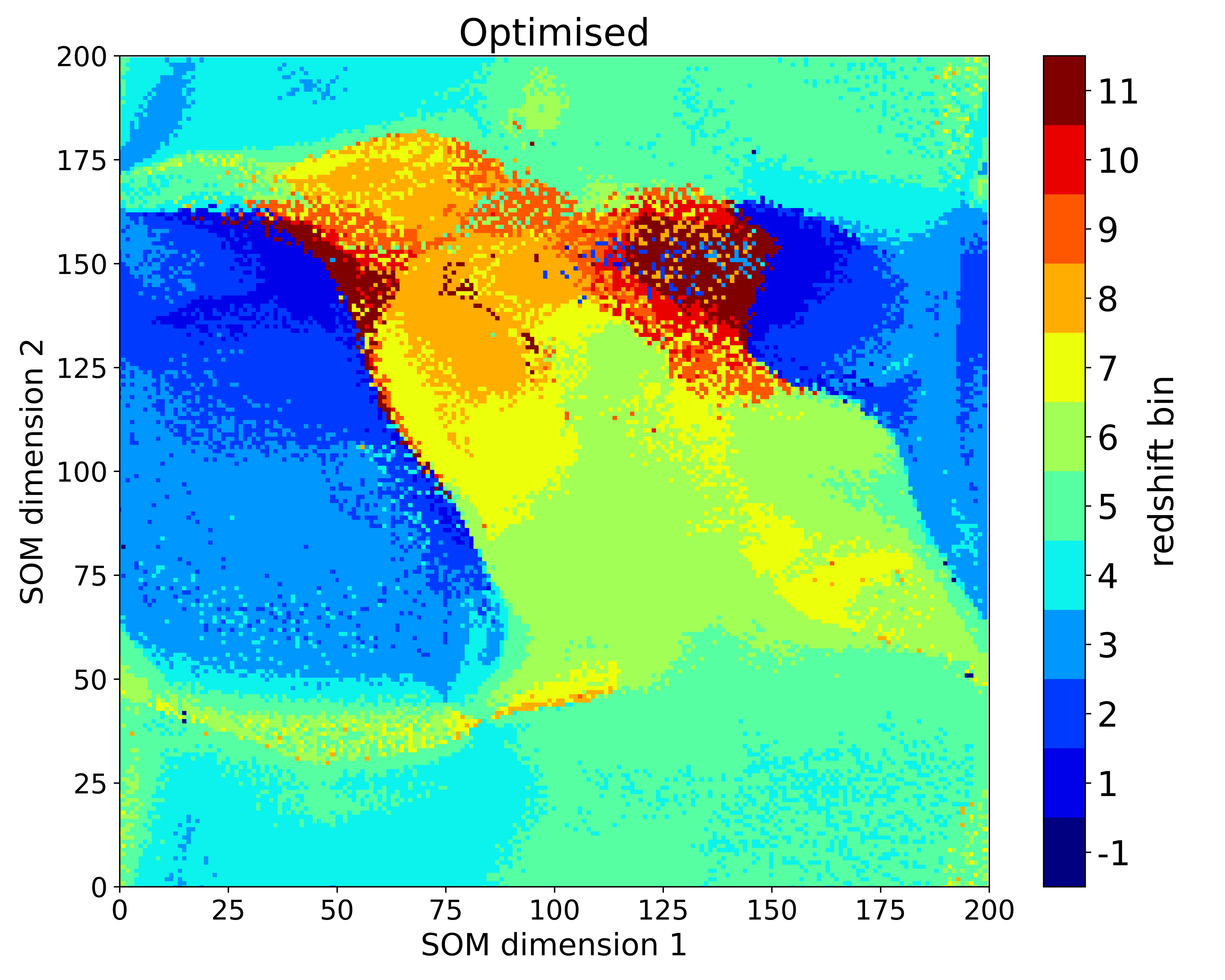} 
        \caption{Colours represent the final redshift bin of each SOM node after running the simulated annealing algorithm.}
	   \label{fig:SOM_final_binned}
    \end{subfigure}
    \hfill
    \begin{subfigure}[t]{0.49\textwidth}
        \centering
        \includegraphics[width=\linewidth]{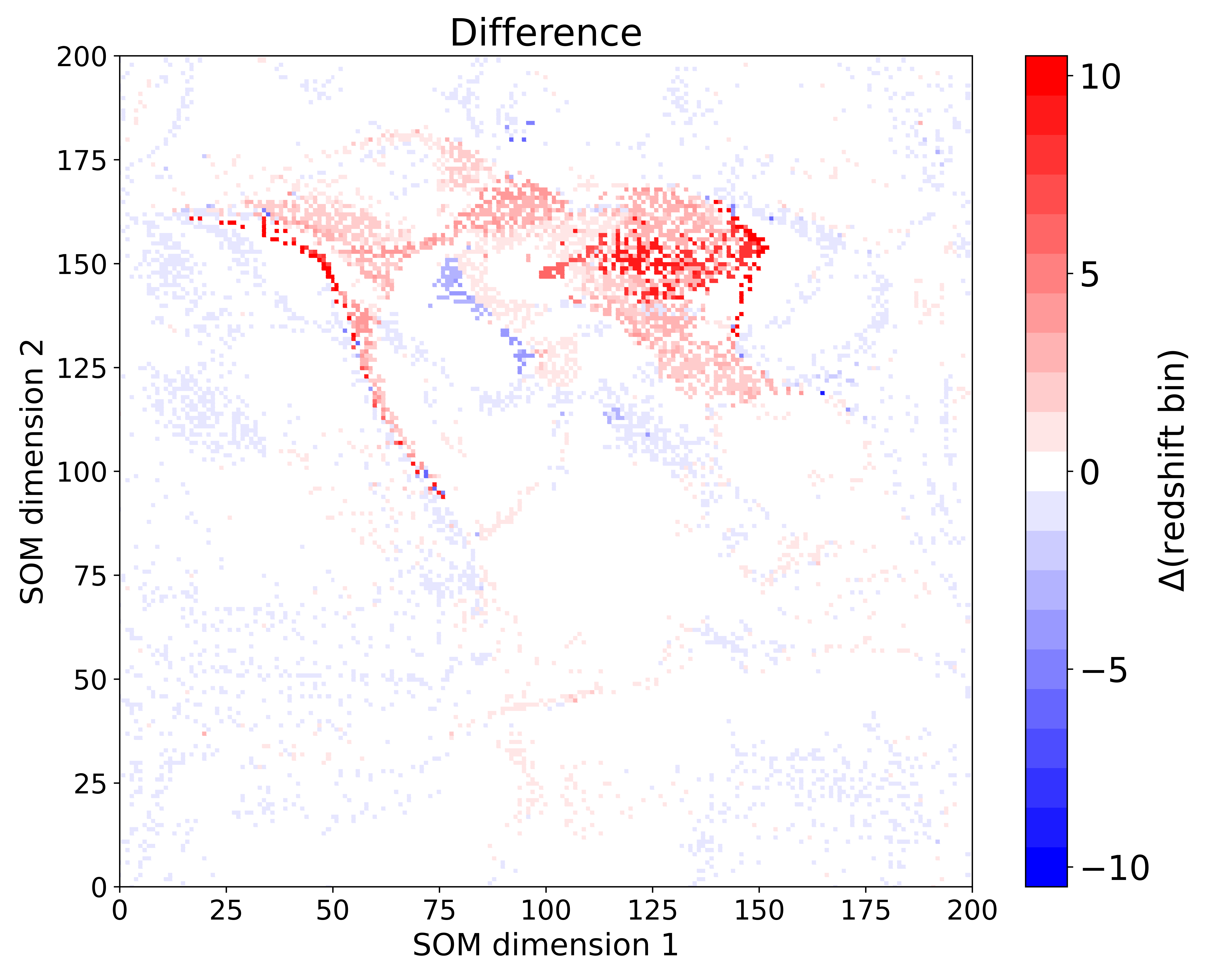} 
        \caption{Colours represent the shift of the redshift bin of each node between panels \subref{fig:SOM_initial_binned} and \subref{fig:SOM_final_binned}.}
	   \label{fig:SOM_changed_binned}
    \end{subfigure}
    \caption{Illustration of the self-organising map used in the analysis. The SOM consists of 200x200 nodes on a rectangular grid in toroidal geometry and is trained on the observed colours of galaxies in the photometric galaxy sample. Coloured labels indicate the tomographic bin to which galaxies in each node are assigned, with SOM nodes which do not contain any galaxies labelled with \enquote*{-1}.}
\end{figure*}

We train a self-organising map and 200x200 nodes on a rectangular grid on the observed colours of galaxies in the photometric sample using the public code {\sc Somoclu} \citep{Wittek13} and choose a toroidal geometry to avoid boundary effects. The SOM is illustrated in Fig. \ref{fig:SOM} with colours representing the mean of the true simulated redshift of galaxies in each node. We note that the true redshift is used solely for illustration purposes and is not used in the further analysis of the photometric sample. In Appendix \ref{ap:SOM} we provide comparisons between SOMs at different resolutions that are inferred from the high-resolution SOM using the method described in Section \ref{sec:SOM}.

As can be observed in Fig. \ref{fig:SOM}, the SOM achieves a separation of galaxies by redshift by relying purely on the colour information of individual galaxies. Furthermore, we find regions where high-redshift nodes are adjacent to low-redshift nodes, which we presume is where catastrophic errors in the photometric redshift estimate preferentially occur \citep{Masters15}. 

\begin{figure*}
    \centering
	\includegraphics[width=\linewidth]{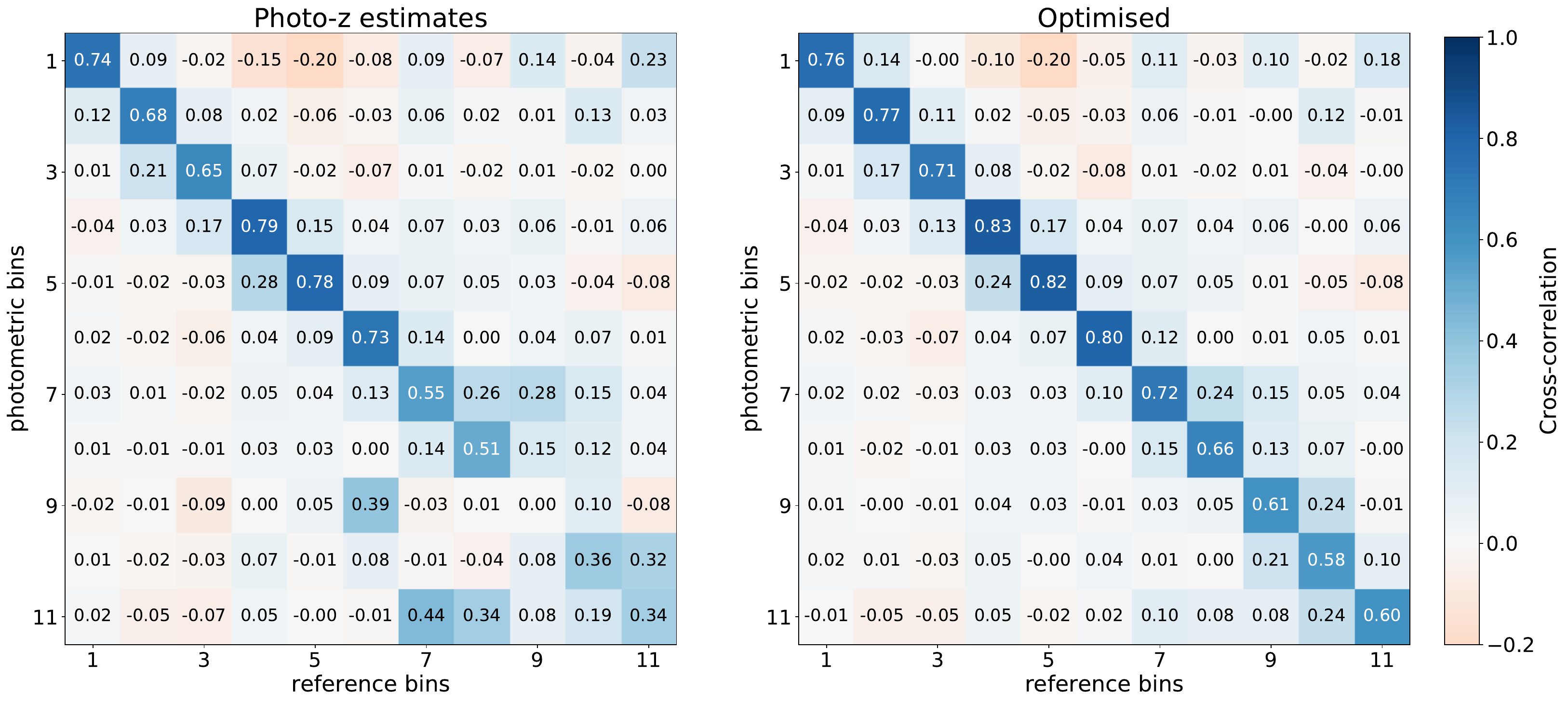}
    \caption{Left: Initial cross-correlation matrix between bins of the photometric sample, inferred from photometric redshift estimates, and the reference sample. Right: Cross-correlation matrix after optimisation of the bin assignment of the photometric sample. A perfect assignment with noise-free clustering measurements would yield the identity matrix.}
    \label{fig:cc_matrix}
\end{figure*}

We assign each SOM node to a redshift bin, depending on the most common photometric redshift bin of galaxies in each node. This is illustrated in Fig. \ref{fig:SOM_initial_binned} with colours indicating tomographic redshift bins. We note that a small number of SOM nodes do not contain any galaxies, so that they cannot be assigned to a redshift bin, which is indicated by a label of \enquote*{-1}. Comparing Fig. \ref{fig:SOM_initial_binned} to the true redshifts, shown in Fig. \ref{fig:SOM}, we find that some SOM nodes show a significant mismatch between the true redshift and photometric redshift bin, especially in the region of high-redshift galaxies. This is further emphasised by the left panel of Fig. \ref{fig:cc_matrix}, which shows the cross-correlation matrix between the photometric and reference samples. We find a rather low correlation signal on the diagonal for high redshift bins, indicating a mismatch of the redshifts of galaxies in these bins. Looking at the off-diagonal elements, we find high non-zero cross-correlation signals between the photometric and reference samples. This suggests that high-redshift bins are contaminated with low redshift galaxies and vice versa, caused by catastrophic failures in the photometric redshift estimation with \textsc{BPZ}.  

We then proceed with re-sorting galaxies in the photometric sample to different tomographic bins using the simulated annealing algorithm described in Section \ref{sec:annealing}. Here, the initial state of the system is the set of redshift bin labels obtained from the photometric redshift estimates illustrated in Fig. \ref{fig:SOM_initial_binned}. We start with an initial SOM resolution of $R_{\rm min} = 30$, obtained with the method outlined in Section \ref{sec:SOM}. The SOM resolution is increased over time until reaching the final resolution of $R_{\rm max} = 80$. This is achieved by coupling the SOM resolution linearly to the temperature of the system which decreases from $T_{\rm max}=1$ to $T_{\rm min}=0.01$. The range of the SOM resolution is chosen so that in the initial phase a larger portion of galaxies is relabelled which then decreases with increasing SOM resolution. We determine the maximum temperature such that initially there is a chance of about 50\% to accept a worse state for a typical value of $\Delta E$ at a resolution of $R_{\min}$. The minimum temperature on the other hand is chosen such that the chance of accepting a worse state at resolution $R_{\rm max}$ approaches zero. The simulated annealing algorithm returns a modified set of redshift bin labels, where SOM nodes were relabelled to different redshifts bins in order to diagonalise the cross-correlation matrix shown in the left panel of Fig. \ref{fig:cc_matrix}. The resulting optimised matrix is shown in the right panel of Fig. \ref{fig:cc_matrix}. We observe that the algorithms succeeds in reducing the cross-correlation signal between photometric and reference bins on the off-diagonal elements while increasing the auto-correlation signal on the diagonal and thus increasing the energy of the system, defined in Eq. \eqref{eq:energy}, from 0.56 to 0.68. In Appendix \ref{ap:energy} we discuss the evolution of the energy during the simulated annealing optimisation.

The resulting redshift bins of each SOM node are illustrated in Fig. \ref{fig:SOM_final_binned}, which is the equivalent to Fig. \ref{fig:SOM_initial_binned}, but instead of the initial redshift bin labels we show the modified labels returned by the algorithm. A comparison of the two figures shows that the simulated annealing algorithm indeed succeeds in identifying those regions of the SOM where the photometric redshift estimates do not match the true redshift of the galaxies and therefore shifts these nodes towards higher redshift bins. This is further illustrated in Fig. \ref{fig:SOM_changed_binned}, where we show the magnitude of the shift in the redshift label per SOM node with positive numbers indicating a shift towards higher redshift bins and negative numbers indicating a shift towards a lower redshift bin. We find that the most significant changes occur in the aforementioned high-redshift nodes and at the boundaries between high- and low-redshift nodes. 

\begin{figure*}
\centering
\includegraphics[width=\linewidth]{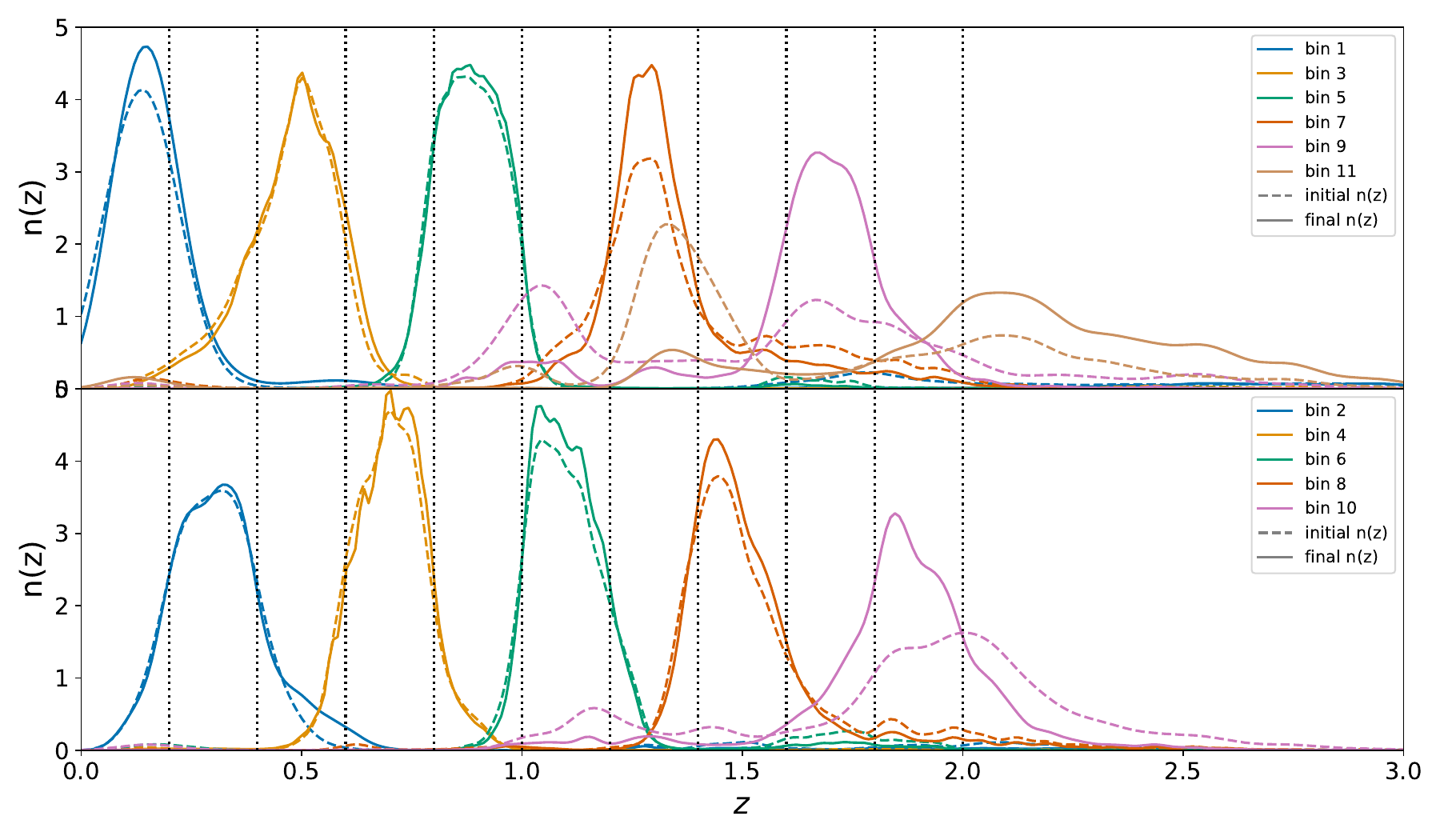}
\caption{Comparison of the initial redshift distribution of each tomographic bin, obtained using the photometric redshift estimate of individual galaxies (dashed lines), and the redshift distribution after simulated annealing (solid lines). Dotted lines indicate the redshift bin edges.}
\label{fig:Nz}
\end{figure*}

The redshift distributions of the tomographic bins are illustrated in Fig. \ref{fig:Nz} with dashed lines indicating the initial redshift distributions and solid lines representing the distributions after relabelling via simulated annealing. We note that the redshift distributions are inferred from the true underlying redshifts of galaxies assigned to each bin, which are not available in a real observational data set and need to be calibrated separately, e.g. via cross-correlation measurements. We find that the algorithm significantly improves the redshift distributions of high-redshift bins, which initially showed significant deviations from the predefined redshift intervals in the tails of the distribution. The correlation matrix shown in Fig. \ref{fig:cc_matrix} indicates a low-level anticorrelation between photometric bin 1 and reference bins 4 and 5 that remains even after optimisation. However, this feature is not observed in Fig. \ref{fig:Nz} which shows no significant overlap between the redshift distribution of bin 1 with either bin 4 or bin 5. The origin of this feature is unclear. 

A possible source of the residual signals are magnification effects causing spurious cross-correlation signals between non-overlapping redshift bins which can lead to biased constraints on cosmological parameters in clustering analyses of upcoming surveys \citep{Lepori22, Mahony22}. The contamination due to magnification was shown to be maximal for bins with low signal-to-noise ratio of galaxy clustering. Thus, the effects of magnification can limit the performance of the optimisation method. We test the impact of magnification on the correlation matrix by measuring the clustering cross-correlation with unmagnified galaxy positions. We find a change in the correlation signal which is smaller than the typical measurement uncertainties. Therefore, we conclude that magnification effects do not significantly limit the performance of the optimisation method employed in this work.

\begin{figure*}
 \centering
    \begin{subfigure}[t]{0.49\textwidth}
        \centering
        \includegraphics[width=\linewidth]{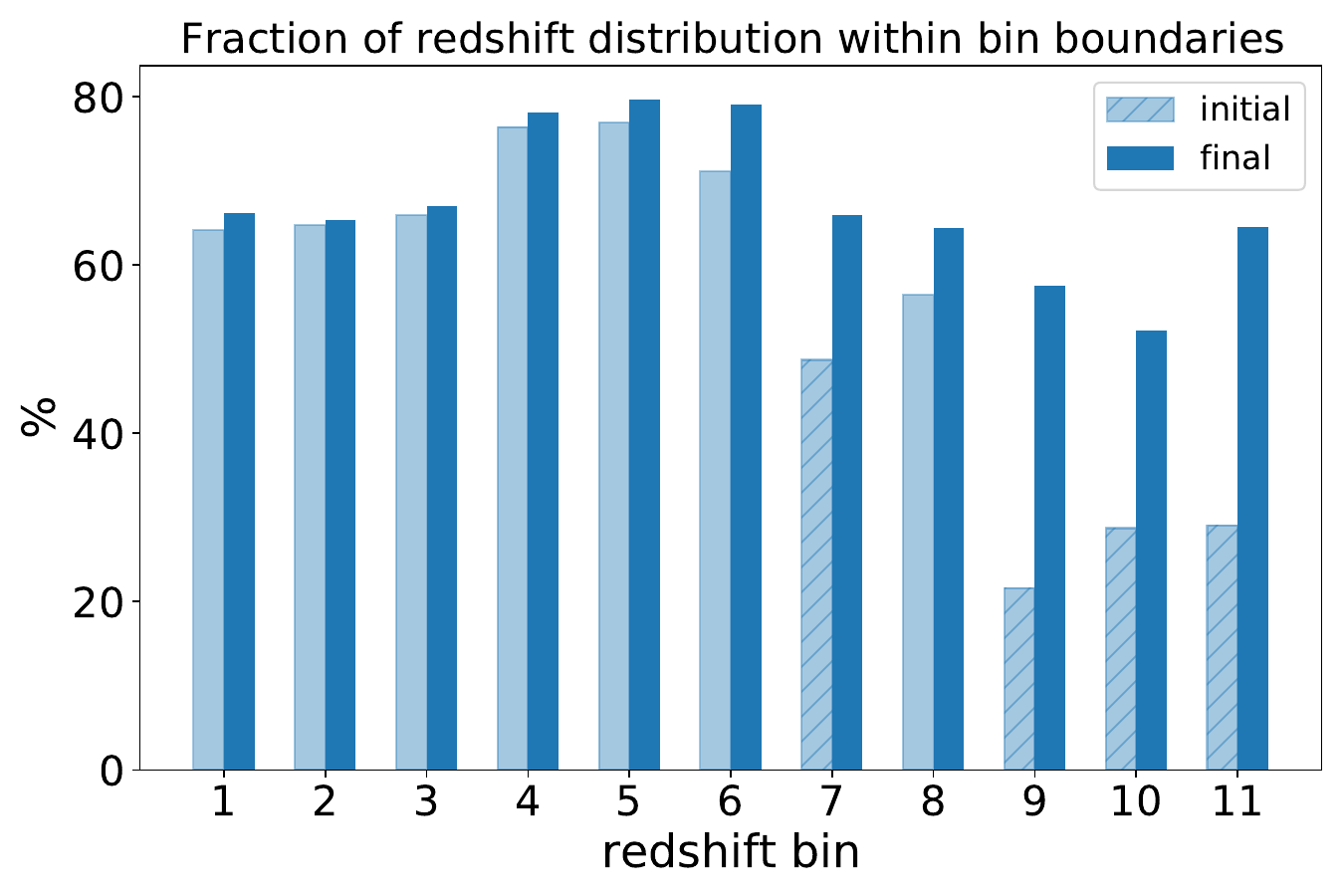} 
    \end{subfigure}
    \hfill
    \begin{subfigure}[t]{0.49\textwidth}
        \centering
        \includegraphics[width=\linewidth]{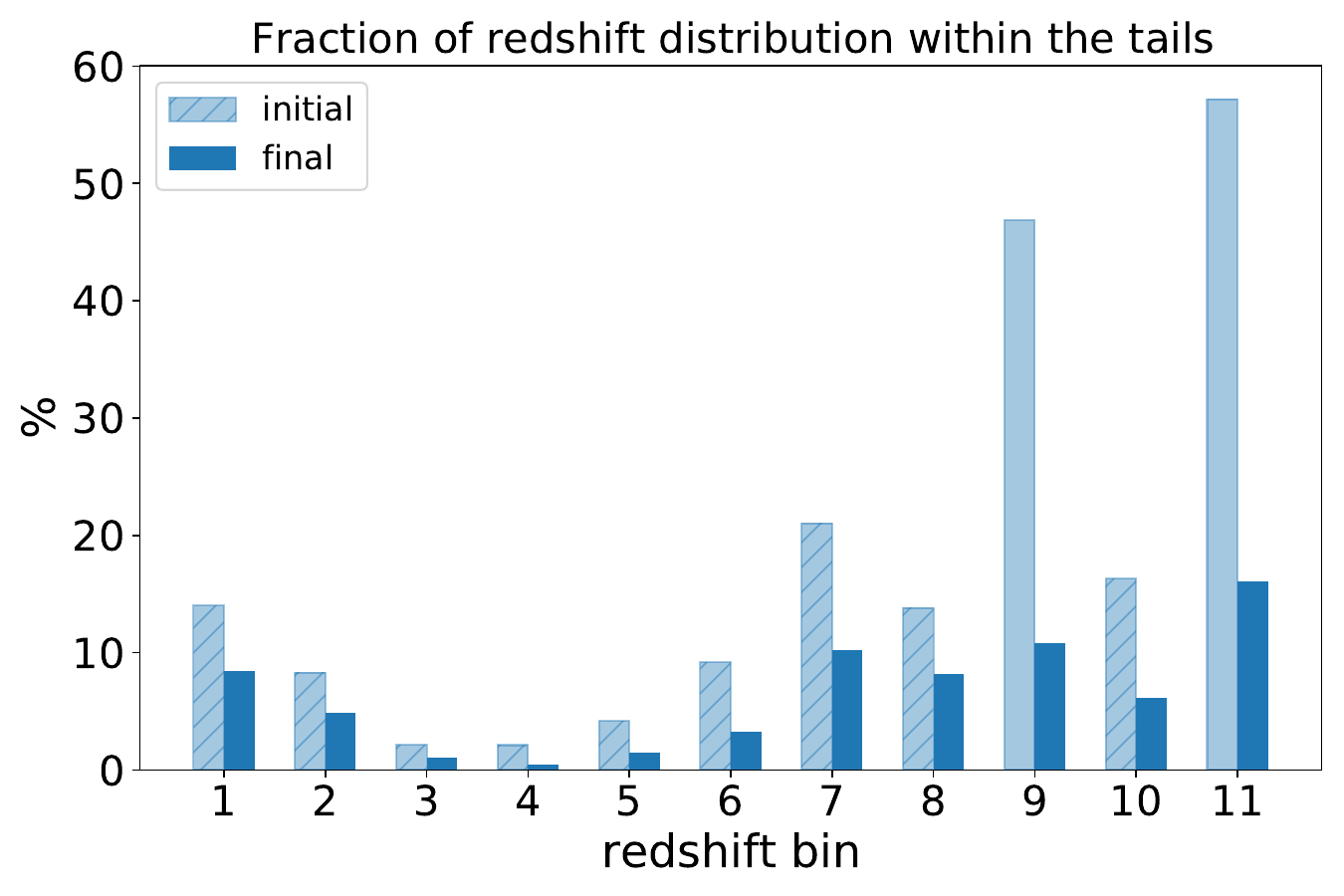} 
	\end{subfigure}
	\caption{Left: Comparison of the percentage of probability mass of the redshift distributions, shown in Fig. \ref{fig:Nz}, within the respective bin range before and after simulated annealing. Right: Comparison of the percentage of the redshift distribution that is located within the tails of the distribution, where we define the tails of the distribution as the region in redshift space that lies more than one bin width outside of the boundaries of a given tomographic bin.}
	\label{fig:percentage}
\end{figure*}

We quantify the extent to which the redshift distributions lie within the boundaries of the redshift bins in the left panel of Fig. \ref{fig:percentage}, comparing the initial and final distribution. We find that the algorithm helps to shift the redshift distribution significantly to lie within the bin boundaries, especially in the higher redshift bins, where we find improvements of about 30 per cent. Additionally, in the right panel of Fig. \ref{fig:percentage} we quantify how much of the redshift distribution is located within the tails of distribution. Here, we define the tails of the distribution as the region in redshift space that lies more than one bin width outside of the boundaries of a given tomographic bin. We find a substantial decrease of the tails of the redshift distribution. Again, the biggest improvements are found in the high redshift bins where initially a large percentage of the distribution is located in the tails, which decreases significantly after the simulated annealing. The biggest change is found in bin 11, which initially only contains 32 per cent of the probability mass within the bin boundaries and a large fraction of about 57 per cent in the tails of the distribution. These quantities shift significantly in the final redshift distribution, with about 64 per cent within the bin boundaries and 16 per cent in the tails. Furthermore, we find significant improvements in bin 9, which initially contains about 22 per cent of galaxies within bin boundaries which increases to 57 per cent. The fraction of galaxies in the tails of this redshift bin also decreases significantly from 47 per cent to 11 per cent.

\section{Conclusions}
\label{sec:conclusions}
In this paper we presented a method, {\sc SharpZ}, to group a sample of galaxies into tomographic redshift bins using estimates of the photometric redshift with subsequent re-sorting using an algorithm that optimises the angular cross-correlation between the photometric galaxy sample and an overlapping sample of reference galaxies. We utilised a simulated annealing algorithm that reassigns groups of galaxies to redshift bins and determines the effect on the cross-correlation matrix by calculating a measure of the diagonality of the matrix. This was combined with a self-organising map (SOM) that was trained on the colour information of photometric galaxies. The SOM allows choosing sets of galaxies that are reassigned in each step of the simulated annealing. Additionally, the resolution of the SOM was increased over time in order to achieve a greater accuracy of the final resulting photometric redshift bins.

We applied this method to a synthetic catalogue, cosmoDC2, that aims to resemble measurements of the upcoming Vera C. Rubin Observatory's Legacy Survey of Space and Time. Our results show that the method significantly reduces the fraction of catastrophic outliers in the tails of the redshift distribution in all tomographic bins, most notably in the highest redshift bins where we find improvements by up to 40 per cent. We found that it succeeds in shifting the redshift distributions towards being within the boundaries of the tomographic bins. High redshift bins show the greatest improvements, where the probability mass within the bin boundaries increases up to about 30 per cent, while the improvement in the low redshift bins, whose redshift distributions initially are already quite compact, is smaller with the probability mass increasing by a few percent. Additionally, we found that the method also greatly reduces the amount of the redshift distribution that is located in the tails of the distribution. Again, we find the biggest improvement in the high redshift bins, where the initial performance of the photometric redshift estimates is worst.

The quality of the redshift distributions inferred with our optimisation method depends on the choice of the SOM parameters. Initially, we trained a SOM consisting of 200 x 200 nodes and scaled the resolution from low to high during the optimisation process. The resolution determines the number of galaxies that are reassigned together to an alternative redshift bin, which scales from large numbers of galaxies to smaller numbers. Thus, the resolution imposes a limit on how well the SOM can separate features in colour space that can be associated with different redshift bins. By choosing a maximum resolution we therefore implicitly put a limit on the quality of the final redshift distribution that can be achieved. The quality of the final redshift distributions can be improved by increasing the maximum resolution at the cost of a longer computation time since a higher resolution implies a selection of fewer galaxies in each step, up to the limit where the algorithm selects individual galaxies. However, at a certain point noise in the clustering observable will limit the observable effect on the objective function. Therefore the method is ultimately limited by both the noise limit of the clustering measurement and the SOM resolution. Additionally, the choice of features on which the SOM is trained influences how well the SOM can separate galaxies of different redshifts. In our analysis we trained the SOM on five galaxy colours, which we found to perform well for the data set considered in this work. Future applications with alternative data sets should however explore alternative sets of training features, for example different colour combinations or the addition of magnitude information, which can help breaking colour-redshift degeneracies.

In our analysis we made the assumption that the reference sample covers the full area observed by the photometric survey, while in a realistic application the reference sample will only have a partial sky overlap with the photometric survey. However, as long as the survey is spatially homogeneous, we can optimise the assignment of galaxies to tomographic bins in the area covered by both the photometric and reference survey and then use the SOM to expand the redshift bin assignments to the full photometric survey. Therefore, a complete overlap of the two samples is not a general requirement of the method. Additionally, the SOM can be used to reproduce the results without re-running the simulated annealing algorithm. Moreover, we can further improve the calculation of the clustering signal by measuring the cross-correlation signal between reference samples and cells of the highest-resolution SOM. For a given assignment of SOM cells to tomographic bins we can then stack the cross-correlation signals of the individual cells using the hierarchical structure of the SOM. In this way, we obtain the correlation signal between photometric and reference bins without re-calculating the cross-correlation in each step of the optimisation algorithm, which will lead to a decrease in computational cost. This is left for future work.

Additionally, we made the assumption that the reference sample is fully representative of the photometric sample. Future applications therefore require studies of how incomplete reference samples and noise influence the cross-correlation measurements between the photometric and reference sample and how it impacts the quality of the inferred tomographic bins. 

While finalising this work, \citet{Zuntz21} put forward a paper on the optimisation of the tomographic binning for the DESC 3x2pt analysis. In particular, the {\sc ComplexSOM} method utilises a matrix of auto- and cross-power spectra, which is a statistic similar to the one employed in this work. While in this work the redshift bin edges of the galaxy sample are fixed, the {\sc ComplexSOM} method instead optimises the parameters that determine the bin edges.

Our work demonstrates that the optimisation method provides a significant improvement of the redshift distribution of a synthetic survey compared to photometric estimates of the redshift. Therefore it provides a promising complement to existing redshift calibration methods in upcoming surveys. An application to observational data is left for future work.

\section*{Acknowledgements}

We thank the anonymous referee for their constructive comments, which helped to improve the manuscript. This work was partially enabled by funding from the UCL Cosmoparticle Initiative. BS acknowledges support from the Max Planck Society and the Alexander von Humboldt Foundation in the framework of the Max Planck-Humboldt Research Award endowed by the Federal Ministry of Education and Research. BS also acknowledges travel support provided by STFC for UK participation in LSST through grant ST/S006206/1. BJ acknowledges support by STFC Consolidated Grant ST/V000780/1. In this analysis we made use of the following software: {\sc NumPy} \citep{numpy}, {\sc SciPy} \citep{scipy}, {\sc Matplotlib} \citep{matplotlib}, {\sc TreeCorr} \citep{Jarvis04}, {\sc BPZ} \citep{Benitez00}, {\sc Somoclu} \citep{Wittek13}, {\sc Astropy} \citep{astropy1, astropy2}, and {\sc Pandas} \citep{pandas}.

This paper has undergone internal review in the LSST Dark Energy Science Collaboration. We would like to thank the internal reviewers Danielle Leonard, Huan Lin, and Bob Armstrong for their constructive comments. We also thank Hendrik Hildebrandt, Angus Wright, and Eric Gawiser for comments on the manuscript.

The DESC acknowledges ongoing support from the Institut National de 
Physique Nucl\'eaire et de Physique des Particules in France; the 
Science \& Technology Facilities Council in the United Kingdom; and the
Department of Energy, the National Science Foundation, and the LSST 
Corporation in the United States.  DESC uses resources of the IN2P3 
Computing Center (CC-IN2P3--Lyon/Villeurbanne - France) funded by the 
Centre National de la Recherche Scientifique; the National Energy 
Research Scientific Computing Center, a DOE Office of Science User 
Facility supported by the Office of Science of the U.S.\ Department of
Energy under Contract No.\ DE-AC02-05CH11231; STFC DiRAC HPC Facilities, 
funded by UK BIS National E-infrastructure capital grants; and the UK 
particle physics grid, supported by the GridPP Collaboration.  This 
work was performed in part under DOE Contract DE-AC02-76SF00515.

For the purpose of open access, the author has applied a Creative Commons Attribution (CC BY) licence to any Author Accepted Manuscript version arising.

Author contributions: BS carried out the analysis and led the writing of the manuscript. BJ and AK provided feedback and suggestions throughout the project and contributed to the writing of the manuscript.

\section*{Data Availability}

The cosmoDC2 catalog is publicly available at: \url{https://portal.nersc.gov/project/lsst/cosmoDC2/_README.html}.



\bibliographystyle{mnras}
\bibliography{references.bib} 




\appendix
\section{Photometric redshifts}
\label{ap:photo-z}
\begin{figure*}
 \centering
    \begin{subfigure}[t]{0.5\textwidth}
        \centering
\includegraphics[width=\linewidth]{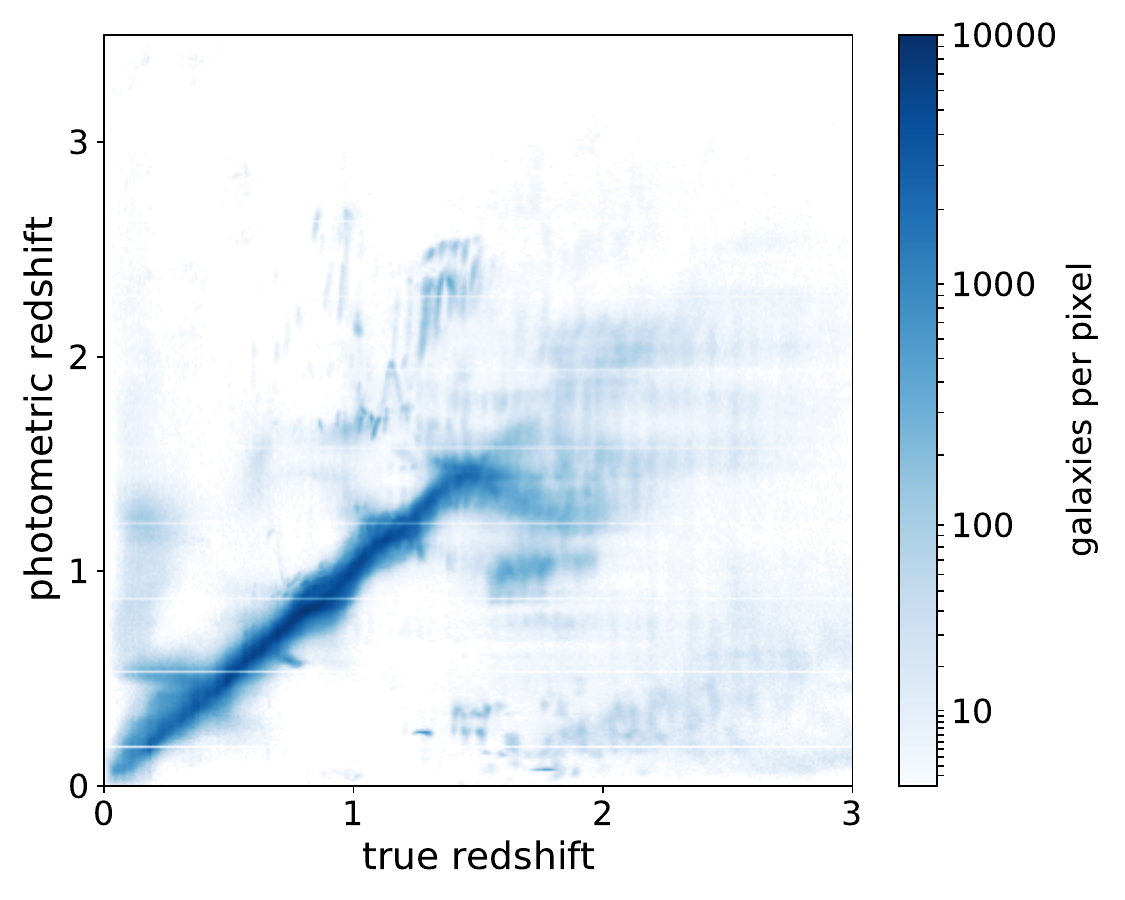}
    \end{subfigure}
    \hfill
    \begin{subfigure}[t]{0.495\textwidth}
        \centering
        \includegraphics[width=\linewidth]{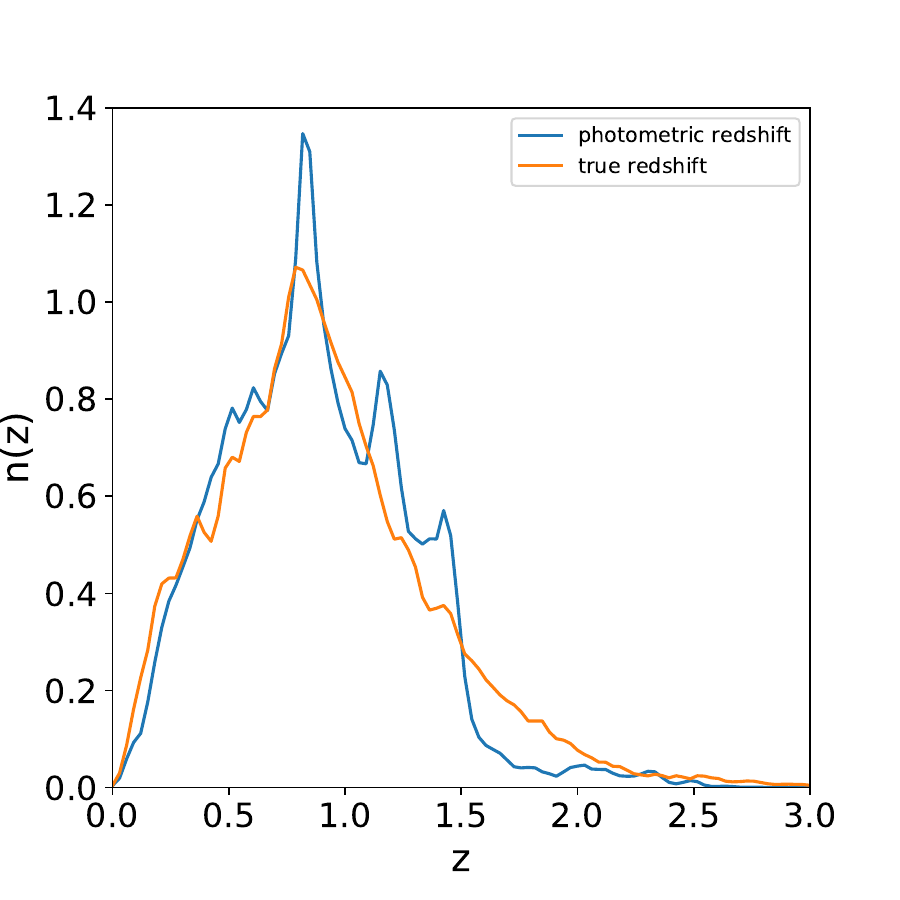} 
	\end{subfigure}
	\caption{Left: Scatter plot of the true redshift of galaxies in the photometric sample and the point estimate of the photometric redshift inferred via SED template fitting. Right: Comparison between the true redshift distribution and the redshift distribution inferred from point estimates of the photometric redshift.}
	\label{fig:phot_vs_spec}
\end{figure*}

In this Appendix we compare the true redshifts of galaxies in the synthetic galaxy catalogue with the point estimate of the photometric redshift inferred via SED template fitting. The left panel of Fig. \ref{fig:phot_vs_spec} shows a scatter plot of the true redshift and the photometric redshift. The redshift distributions inferred from the true redshift and the photometric redshift, respectively, are illustrated in the right panel of Fig. \ref{fig:phot_vs_spec}.
\section{SOM clustering}
\label{ap:SOM}
\begin{figure*}
\centering
\includegraphics[width=\linewidth]{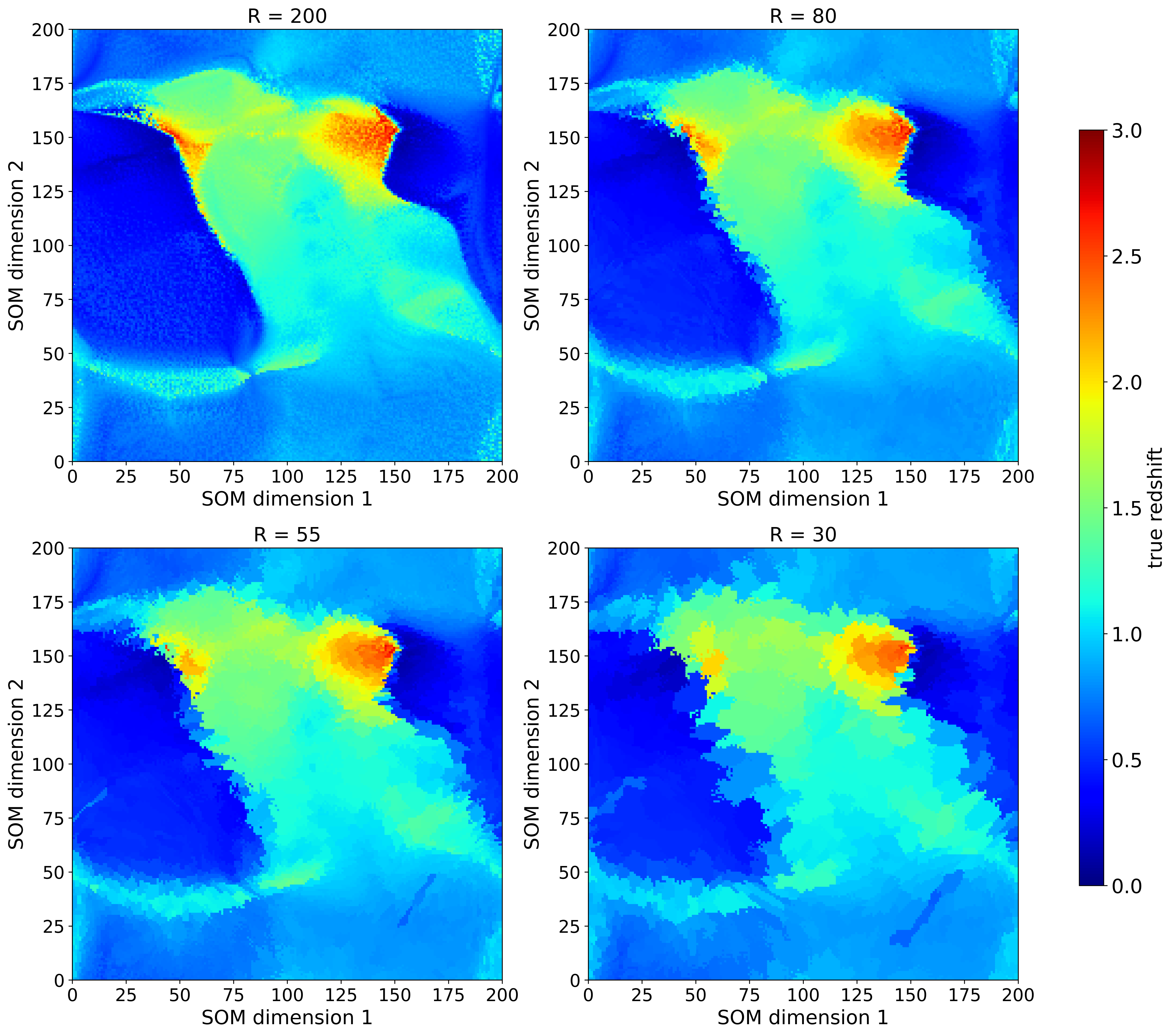}
\caption{Illustration of the SOM used in the analysis at different resolutions. The original SOM, trained on the observed colours of galaxies with a resolution of $R = 200$, is illustrated in the top left panel. The remaining panels show the SOM with reduced resolution inferred with the clustering method described in Sect. \ref{sec:SOM}. The colours represent the mean of the true redshift of galaxies in each node.}
\label{fig:som_res}
\end{figure*}

In this Appendix we provide a comparison of SOMs with decreased resolutions derived from a high-dimensional SOM using the hierarchical clustering method described in Section \ref{sec:SOM}. The original SOM, trained on the observed colours of galaxies with a resolution of $R = 200$, is illustrated in the top left panel of Fig. \ref{fig:som_res}. The remaining panels show SOMs with lower resolutions, inferred from the original SOM via clustering of the weight vectors. The bottom right panel shows the SOM with a resolution of $R=30$, which is the initial resolution from which the simulated annealing algorithm selects groups of galaxies. The top right panel shows the SOM with the final resolution $R=80$, while a SOM with an intermediate resolution of $R=55$ is illustrated in the bottom left panel. The colours in each panel represent the mean of the true redshift of galaxies in each node. We note that since the low-resolution SOMs are constructed from the high-resolution SOM with $R=200$, the axes in each panel refer to the index of the high-dimensional SOM.
\section{Energy in the simulated annealing optimisation}
\label{ap:energy}
\begin{figure}
\centering
\includegraphics[width=\linewidth]{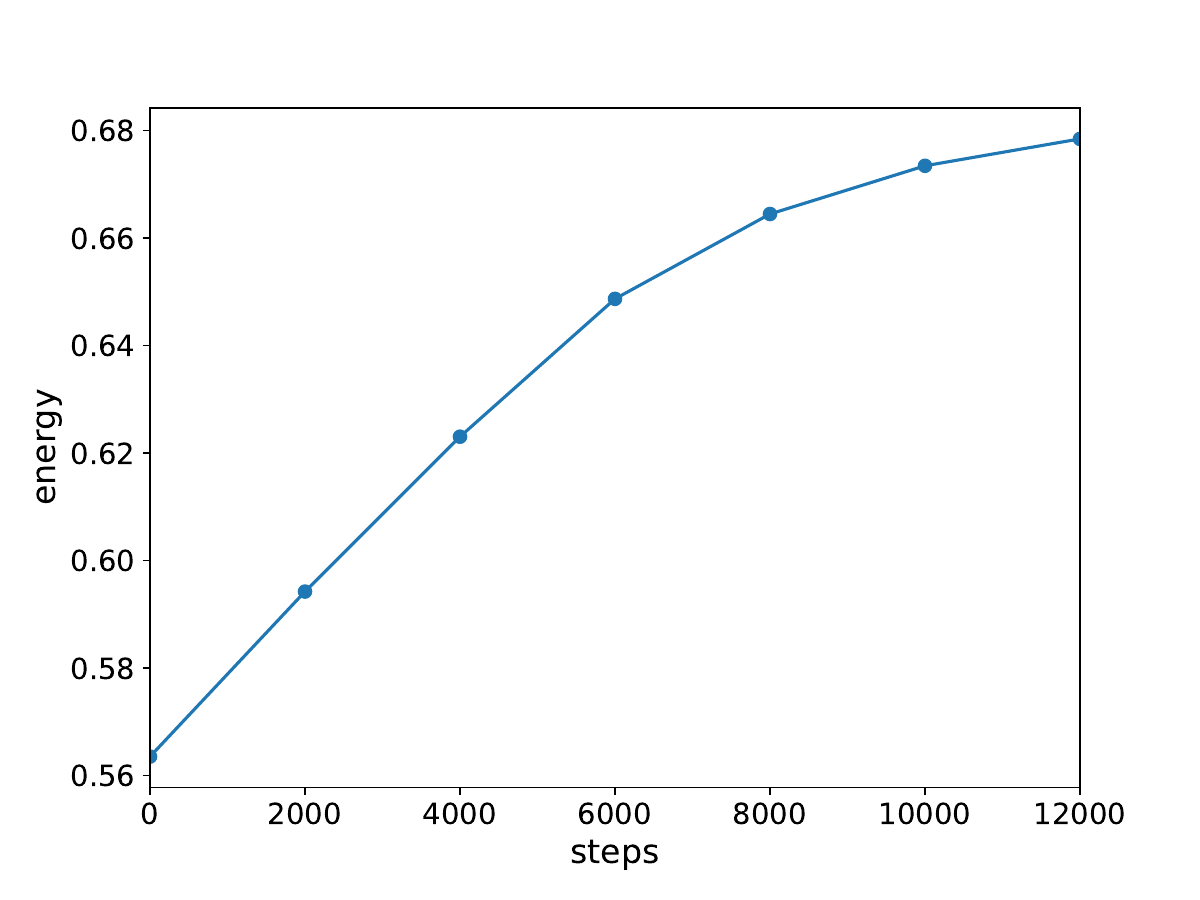}
\caption{Evolution of the energy of the simulated annealing algorithm.}
\label{fig:energy}
\end{figure}

In this appendix we show the evolution of the energy during the simulated annealing optimisation. In Fig. \ref{fig:energy} we illustrate the energy of the system after six iterations of the algorithm. Each iteration corresponds to a full run of the algorithm with an initial SOM resolution of $R_{\rm min} = 30$ and a final resolution $R_{\rm max}=80$ with $N_{\rm steps} = 2000$ \footnote{The computing time for one step on a 16 core machine is approximately 10 seconds}. We observe that in the first three iterations the algorithm achieves an approximately equal increase in the energy of the system, while the later iterations show smaller increases in the energy, indicating that the algorithm converges towards the maximum energy. The method of consecutively running the algorithm multiple times allows us to explore how many steps in total are needed for the algorithm to converge towards the maximum achievable energy for a given final resolution $R_{\rm max}$. We note that after obtaining the final assignment of galaxies to tomographic bins via simulated annealing, this result can potentially be further improved by re-running the algorithm with an initial resolution of $R_{\rm min} = 80$ and an even higher resolution $R_{\max}$ which can be increased up to the initial resolution of the SOM. However, this comes at the cost of a longer runtime, since higher resolutions imply a selection of fewer galaxies in each step, up to the limit where the algorithm selects individual galaxies. Furthermore, at a certain point noise in the clustering observable will limit the observable effect on the energy of the system.

\bsp	
\label{lastpage}
\end{document}